\documentclass[conference]{IEEEtran}

\usepackage{graphicx}
\usepackage{amsmath}
\usepackage{hyperref}
\usepackage{booktabs}
\usepackage{float}
\usepackage{cite}
\usepackage{tikz}
\usetikzlibrary{positioning, arrows.meta}

\title{Streaming REST APIs for Large Financial Transaction Exports from Relational Databases}

\author{
\IEEEauthorblockN{Abhiram Kandiraju}
\IEEEauthorblockA{
Distinguished Engineer, Capital One\\
Richmond, Virginia, USA
}
}

\begin{document}

\maketitle

\noindent\textbf{ABSTRACT}\\
Financial platforms and enterprise systems frequently provide transaction export capabilities to support reporting, reconciliation, auditing, and regulatory compliance workflows. In many environments, these exports involve very large datasets containing hundreds of thousands or even millions of transaction records. Traditional REST API implementations often construct the entire export payload in application memory before transmitting the response to the client, which can lead to high memory consumption and delayed response initiation when processing large datasets.

This paper presents a streaming-based REST API architecture that retrieves transaction records incrementally from relational databases and writes them directly to the HTTP response output stream. By integrating database cursor retrieval with progressive HTTP transmission, the proposed design allows export data to be delivered continuously as records are processed rather than after the full dataset has been assembled.

The architecture is implemented using a Java-based JAX-RS framework with the \texttt{StreamingOutput} interface and supports multiple financial export formats including CSV, OFX, QFX, and QBO. In practice, the streaming approach significantly reduces memory buffering requirements and allows large export downloads to begin immediately, improving responsiveness and scalability for high-volume export operations.

\noindent\textbf\\
{KEYWORDS --} Streaming APIs, REST Architecture, Data Export, Financial Systems

\section{Introduction}

Modern financial platforms commonly provide transaction history export capabilities to support accounting reconciliation, regulatory compliance, auditing, and tax reporting workflows. These exports are typically delivered through REST-based APIs that allow client applications or web browsers to retrieve transaction records over HTTP-based interfaces. While this approach is straightforward to implement, delivering very large transaction datasets efficiently presents several architectural challenges.

In many traditional implementations, the application server retrieves the full result set from the database, constructs the entire export payload in memory, and only then transmits the response to the client. Although this design simplifies implementation, it can become inefficient when the dataset is large. Generating large response payloads entirely in memory increases memory consumption on the application server and delays the initiation of the HTTP response. From the user's perspective, the request may appear idle even though the system is actively processing a substantial amount of data.

These limitations become more pronounced in enterprise financial systems where export requests may involve hundreds of thousands or even millions of transaction records. When multiple users request exports concurrently, large in-memory payload generation can lead to increased garbage collection activity, longer response times, and reduced system scalability.

A more efficient approach is to treat the export process as a continuous data stream rather than as a single aggregated payload. Instead of constructing the entire export file before transmission begins, records can be retrieved incrementally from the database and written directly to the HTTP response output stream. This streaming approach allows the server to begin transmitting data immediately while continuing to retrieve and serialize additional records.

Efficient processing of large result sets has long been an important concern in database systems, particularly in the context of transaction processing and large-scale data retrieval strategies \cite{bernstein}. This paper presents a practical streaming architecture for implementing large transaction export services in REST-based systems. The proposed design integrates incremental database cursor retrieval with HTTP response streaming so that transaction records can be delivered progressively as they are read from the database. The implementation is demonstrated using a Java-based JAX-RS framework with the \texttt{StreamingOutput} interface, enabling efficient delivery of large export datasets without requiring full response buffering.

\noindent This paper makes the following contributions:

\begin{itemize}
	\item It analyzes the scalability and latency limitations of traditional buffered export mechanisms commonly used in REST-based transaction APIs.
	
	\item It proposes a streaming-based export architecture that integrates incremental database retrieval with HTTP response streaming to enable efficient delivery of large transaction datasets.
	
	\item It presents a practical implementation using the JAX-RS \texttt{StreamingOutput} interface that supports multiple financial export formats while maintaining a stateless REST service model.
	
	\item It discusses operational considerations and scalability implications for deploying streaming export pipelines in enterprise financial systems.
\end{itemize}

The remainder of this paper is organized as follows. Section II provides background on conventional REST-based data delivery approaches. Section III discusses the limitations of traditional buffered export implementations. Section IV presents the proposed streaming architecture. Section V describes the implementation details of the streaming export mechanism. Section VI explains support for multiple financial export formats. Section VII discusses performance implications, and Section VIII outlines operational considerations for production deployments. Section IX concludes the paper.

\section{Background}

Exporting large datasets is a common requirement in enterprise financial systems. Modern distributed applications frequently expose data and services through RESTful APIs because of their simplicity and compatibility with standard HTTP-based web infrastructure \cite{pautasso}. Efficient delivery of large datasets and scalable service architectures has been widely studied in distributed systems and web services literature \cite{tanenbaum,armbrust}. In financial platforms, export workflows typically retrieve large record sets from transactional databases and serialize the results into formats such as CSV or financial interchange standards.

In conventional architectures, the export process generally follows the steps below:

\begin{enumerate}
	\item The API retrieves the entire dataset from the database.
	\item The application serializes the full dataset into the desired output format.
	\item The completed response payload is returned to the client.
\end{enumerate}

Although this approach is straightforward to implement, it becomes inefficient when export datasets grow large. Constructing the entire response payload in memory increases memory consumption on the application server and delays the initiation of the HTTP response. As dataset sizes grow, these limitations can lead to higher latency, increased garbage collection activity, and reduced scalability when multiple export requests are processed concurrently.

While HTTP response streaming and database cursor-based retrieval are well-known techniques in web and database systems, their combined use for large-scale financial export services introduces several practical challenges. These include maintaining stateless REST interactions, supporting multiple financial interchange formats, and operating within production constraints such as browser download behavior, connection pool limits, and long-running export requests.

The contribution of this work is therefore not the introduction of new underlying protocols, but the presentation of a practical architecture that integrates these established techniques into a deployable pattern for high-volume financial transaction exports. The design demonstrates how incremental database retrieval, format-specific serialization, and HTTP streaming can be combined to produce scalable export services capable of delivering very large datasets without excessive memory buffering.

\subsection{Related Work}
HTTP streaming and chunked transfer mechanisms have long been used to deliver large responses incrementally in web services. Similarly, database cursor mechanisms and streaming result sets are commonly used to process large query results without materializing full datasets in application memory.

More advanced streaming systems may incorporate backpressure-aware processing pipelines or reactive data flows. However, such frameworks often introduce additional architectural complexity that may not be necessary for many enterprise export workflows. The approach presented in this paper focuses on a simpler pattern that can be implemented using widely available REST frameworks and relational database access mechanisms while still supporting large-scale transaction exports.

\section{Limitations of Conventional Export Implementations}

Traditional export implementations typically follow a buffered response model in which the application retrieves the entire dataset from the database, constructs the export payload in memory, and only then transmits the response to the client. While this design simplifies application logic, it introduces several performance and scalability limitations when export datasets become large.

First, buffering large datasets in application memory can significantly increase memory consumption on the application server. When export requests involve hundreds of thousands or millions of records, the server must allocate memory for both the retrieved dataset and the serialized output representation. This can lead to increased garbage collection activity and potential memory pressure under heavy workloads.

Second, response latency increases because the server cannot begin transmitting the export file until the entire dataset has been retrieved and serialized. From the user's perspective, the request may appear inactive even though the system is actively processing the data.

Third, buffered export approaches reduce overall system scalability. When multiple large export requests are processed concurrently, each request may consume substantial memory and CPU resources. This can limit the number of concurrent export operations the system can support and may negatively affect the performance of other services running on the same infrastructure.

These limitations highlight the need for an alternative export architecture that can handle large transaction datasets more efficiently while maintaining acceptable response times and system scalability.

\section{Streaming Architecture}

Traditional export workflows in REST-based systems commonly follow a buffered response model. In this approach, the application retrieves the full dataset from the database, constructs the complete export payload in memory, serializes it into the desired output format, and only then returns the response to the client. While this design is straightforward to implement, it becomes inefficient when the export dataset is large.

\begin{figure}[!ht]
	\centering
	\begin{tikzpicture}[
		node distance=1.2cm,
		box/.style={
			rectangle,
			draw,
			rounded corners,
			fill=gray!10,
			minimum width=0.8in,
			minimum height=0.28in,
			align=center,
			font=\scriptsize
		},
		arrow/.style={->, line width=0.8pt}
		]
		
		\node[box] (db) {Database};
		\node[box, right=of db] (buffer) {Memory\\Buffer};
		\node[box, right=of buffer] (serialize) {Serialize};
		
		\node[box, below=0.55cm of serialize] (http) {HTTP\\Response};
		\node[box, left=of http] (wait) {Wait for\\Complete File};
		\node[box, left=of wait] (client) {Client\\Download};
		
		\draw[arrow] (db) -- (buffer);
		\draw[arrow] (buffer) -- (serialize);
		\draw[arrow] (serialize) -- (http);
		\draw[arrow] (http) -- (wait);
		\draw[arrow] (wait) -- (client);
		
	\end{tikzpicture}
	\caption{Traditional buffered export architecture illustrating full payload buffering before transmission.}
	\label{fig:buffered_architecture}
\end{figure}
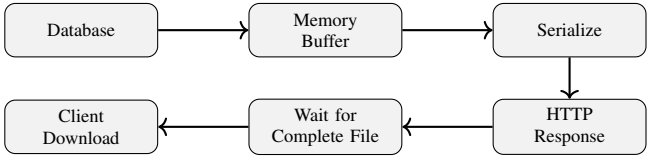

Figure \ref{fig:buffered_architecture} illustrates the conventional buffered export workflow.

To address these limitations, a streaming-based architecture can be used to deliver large datasets incrementally as they are retrieved from the database. Instead of constructing the entire export payload in memory before sending a response, the server progressively retrieves records, transforms them into the required export format, and writes them directly to the HTTP response output stream.

In this architecture, database retrieval, record transformation, encoding, serialization, and HTTP transmission operate as a continuous pipeline. As records are fetched from the database, they are immediately processed and transmitted to the client. This allows the server to begin sending data as soon as the first records become available, significantly reducing perceived response latency.

The streaming pipeline can be conceptually divided into six stages:

\begin{enumerate}
	\item \textbf{DB Fetch} --- Transaction records are retrieved incrementally from the relational database.
	\item \textbf{Cursor} --- A forward-only cursor or result set iterator is used to process records sequentially.
	\item \textbf{Record Map} --- Each database row is mapped to a transaction export record containing the required fields.
	\item \textbf{Encode} --- The mapped record is prepared according to the selected export format.
	\item \textbf{Serialize} --- The encoded record is converted into the final output representation.
	\item \textbf{HTTP Stream} --- The serialized output is written directly to the HTTP response stream and transmitted progressively to the client.
\end{enumerate}

\begin{figure}[!ht]
	\centering
	\begin{tikzpicture}[
		node distance=1.2cm,
		box/.style={
			rectangle,
			draw,
			rounded corners,
			fill=gray!10,
			minimum width=0.8in,
			minimum height=0.28in,
			align=center,
			font=\scriptsize
		},
		arrow/.style={->, line width=0.8pt}
		]
		
		\node[box] (db) {DB\\Fetch};
		\node[box, right=of db] (cursor) {Cursor};
		\node[box, right=of cursor] (map) {Record\\Map};
		
		\node[box, below=0.55cm of map] (encode) {Encode};
		\node[box, left=of encode] (serialize) {Serialize};
		\node[box, left=of serialize] (stream) {HTTP\\Stream};
		
		\draw[arrow] (db) -- (cursor);
		\draw[arrow] (cursor) -- (map);
		\draw[arrow] (map) -- (encode);
		\draw[arrow] (encode) -- (serialize);
		\draw[arrow] (serialize) -- (stream);
		
	\end{tikzpicture}
	\caption{Streaming export pipeline architecture showing incremental retrieval and progressive HTTP transmission.}
	\label{fig:streaming_architecture}
\end{figure}
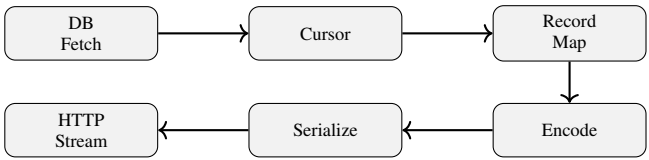

Figure \ref{fig:streaming_architecture} illustrates the proposed streaming export pipeline architecture.

This architecture significantly reduces memory consumption because the application no longer needs to retain the entire export dataset in memory. Instead, only a small portion of the dataset is processed at any given time. As a result, the server can handle large exports more efficiently while maintaining the stateless interaction model of the REST service.

Another important advantage of this approach is improved responsiveness. Because the response stream begins as soon as the first records are available, the client can start receiving data immediately rather than waiting for the complete export file to be assembled. This is particularly valuable in browser-based download workflows, where long delays before response initiation can create the impression that the request is inactive.

By aligning incremental database retrieval with progressive HTTP transmission, the streaming architecture provides a practical and scalable design for large transaction export services in enterprise financial systems. Although export requests may remain open for extended periods while large datasets are transmitted, the service maintains a stateless interaction model. All filtering parameters and export options are provided as part of the HTTP request, and the server does not maintain session-specific state beyond the lifetime of the request itself.

\section{Implementation}

The streaming export architecture can be implemented using standard web service frameworks that support incremental response writing. In this implementation, the export service is built using the JAX-RS\cite{jaxrs} framework and the \texttt{StreamingOutput} interface, which allows the application to write data directly to the HTTP response output stream as records are processed.

When a client initiates an export request, the API endpoint first validates the request parameters, including filters, date ranges, and the export format requested by the client. After validation, the service prepares the database query that retrieves the required transaction records. Rather than loading the entire dataset into application memory, the query is executed using a forward-only cursor or streaming result set so that records can be processed sequentially. In practical implementations, database cursor behavior must be configured carefully to ensure efficient streaming retrieval. For example, JDBC-based implementations may rely on forward-only result sets and configurable fetch sizes so that rows are retrieved incrementally from the database rather than fully materialized in application memory. These settings allow the export service to process large datasets while maintaining stable memory usage. For very large exports, it is also important to consider transaction isolation behavior and snapshot semantics so that long-running export queries do not interfere with concurrent transactional workloads.

Once the query execution begins, the application enters a streaming processing loop. Each record retrieved from the database is immediately mapped to an intermediate export record structure containing the fields required by the export format. The mapped record is then encoded and serialized into the selected output format before being written to the HTTP response stream.

The core processing workflow can be summarized as follows:

\begin{enumerate}
	\item Validate the export request parameters and determine the requested export format.
	\item Execute the database query using a prepared statement configured for streaming retrieval.
	\item Iterate through the result set using a forward-only cursor.
	\item Transform each database row into a transaction export record.
	\item Encode the record according to the selected export format (e.g., CSV, OFX, QFX, or QBO).
	\item Serialize the encoded record into the output representation.
	\item Write the serialized data directly to the HTTP response output stream.
	\item Continue processing until all records have been retrieved and transmitted.
\end{enumerate}

Because the response stream is written incrementally, the client begins receiving the export file immediately after the first records are processed. This significantly reduces response initiation latency compared with buffered export implementations, where the client must wait until the entire dataset has been assembled before receiving any data.

The streaming implementation also improves resource utilization on the application server. At any given time, only a small portion of the dataset is held in memory, typically limited to the current record being processed and a small serialization buffer. This allows the service to support large export requests while maintaining stable memory usage even under concurrent workloads.

Error handling is an important consideration in streaming implementations. Input validation errors can be detected before streaming begins and returned using standard HTTP error responses. However, once the response stream has started, HTTP headers have already been sent to the client and conventional error responses are no longer possible. In such cases, the service must terminate the stream gracefully if an exception occurs during processing.

This implementation approach enables efficient generation of large transaction export files while maintaining a stateless REST service model and minimizing memory overhead on the application server.

\section{Multi-Format Export Support}

Enterprise financial platforms frequently require transaction exports in multiple industry-standard formats so that the exported data can be imported into different accounting and financial management systems. To support this requirement, the streaming export architecture separates the data retrieval pipeline from the format-specific serialization logic.

In the proposed design, transaction records retrieved from the database are first mapped to a common intermediate representation. This representation contains the normalized fields required for export, such as transaction identifiers, timestamps, account references, and monetary values. By introducing this intermediate structure, the system can apply different format encoders without modifying the core data retrieval pipeline.

Each export format is implemented as a dedicated serializer component responsible for converting the intermediate transaction record into the required output format. The architecture supports several commonly used financial export formats, including:

\begin{itemize}
	\item \textbf{CSV} --- a widely used tabular format that can be imported into spreadsheets and data analysis tools.
	\item \textbf{OFX} --- the Open Financial Exchange format commonly used by financial institutions for transaction data exchange.
	\item \textbf{QFX} --- a variant of OFX used by certain financial software applications.
	\item \textbf{QBO} --- a format compatible with QuickBooks financial management systems.
\end{itemize}

Because serialization occurs as part of the streaming pipeline, each record is encoded and written to the response stream as soon as it is processed. This allows large export files to be generated progressively without requiring format-specific buffering of the entire dataset.

This modular format-handling approach improves extensibility, since additional export formats can be introduced by implementing new serializer components without modifying the underlying streaming architecture.

\section{Performance Discussion}

Streaming architectures provide several operational advantages compared to traditional buffered export implementations. The most significant improvement comes from reducing the need to construct large response payloads in application memory.

In large-scale computing systems, efficient use of memory resources is critical for maintaining predictable system performance under concurrent workloads \cite{patterson}. In buffered export workflows, the application server must allocate memory to hold the entire dataset while it is being serialized. For large transaction exports, this can involve hundreds of thousands or even millions of records. When multiple export requests are processed concurrently, memory consumption can increase rapidly, potentially leading to increased garbage collection activity and reduced system stability.

By contrast, the streaming architecture processes records incrementally. Each record is retrieved from the database, transformed, serialized, and written directly to the HTTP response stream. Because only a small portion of the dataset is processed at any given time, the memory footprint of the export operation remains relatively constant regardless of the total dataset size.

Another important benefit is improved response initiation time. In buffered implementations, the client must wait until the entire dataset has been retrieved and serialized before the response can begin. With streaming, the server can start transmitting data as soon as the first records become available. This reduces perceived latency and improves the responsiveness of export operations, particularly for browser-based downloads.

Streaming architectures also improve system scalability. Since export operations no longer require large in-memory buffers, the application server can handle more concurrent export requests without exhausting available resources. This is especially valuable in enterprise environments where multiple users may request transaction exports simultaneously.

Table~\ref{comparison} summarizes the qualitative differences between traditional buffered export approaches and streaming-based export implementations.

\begin{table}[!ht]
	\centering
	\caption{Traditional API vs Streaming API}
	\label{comparison}
	\begin{tabular}{lcc}
		\hline
		Metric & Traditional API & Streaming API \\
		\hline
		Memory Usage & High & Low \\
		Response Start Time & Delayed & Immediate \\
		Large Dataset Support & Limited & High \\
		Scalability & Moderate & High \\
		\hline
	\end{tabular}
\end{table}

Overall, the streaming architecture provides a practical approach for improving the efficiency and scalability of large transaction export services in REST-based enterprise systems.

\subsection{Illustrative Evaluation}
To illustrate the operational differences between buffered and streaming export approaches, a representative synthetic export workload was executed using datasets ranging from 100,000 to 1,000,000 transaction records. The results demonstrate the expected characteristics of the streaming architecture: significantly lower peak memory usage and faster response initiation.

\begin{table}[!ht]
	\centering
	\caption{Illustrative export behavior}
	\begin{tabular}{lccc}
		\hline
		Dataset Size & Approach & Time to First Byte & Peak Memory \\
		\hline
		100k rows & Buffered & High & High \\
		100k rows & Streaming & Immediate & Low \\
		1M rows & Buffered & Delayed & High \\
		1M rows & Streaming & Immediate & Low \\
		\hline
	\end{tabular}
\end{table}

\section{Operational Considerations and Scalability}

While the streaming export architecture significantly improves memory efficiency and response latency, several operational considerations must be addressed when deploying such a system in production environments.

One important factor is database query behavior. Large export requests may involve retrieving hundreds of thousands or even millions of transaction records. Although the streaming approach avoids loading the entire dataset into application memory, the underlying database query must still be optimized to ensure stable performance. This typically involves appropriate indexing on transaction tables and the use of efficient query filters such as date ranges or account identifiers.

Another consideration is connection management. Because the export stream remains active while records are retrieved and transmitted, the database connection and HTTP response remain open for the duration of the export operation. In systems with high concurrency, connection pooling strategies must be carefully configured to ensure that long-running export operations do not exhaust available database connections.

Network behavior also plays an important role in streaming-based exports. Since the response is transmitted incrementally, the server sends the export data in multiple chunks rather than as a single large payload. This approach reduces memory pressure on the server but may result in longer-lived HTTP connections. Proper timeout configurations and buffering policies are therefore important to ensure reliable delivery.

From a client perspective, the streaming model improves perceived responsiveness because the browser begins receiving the export file almost immediately after the request is initiated. Users can observe the download progressing while the server continues generating the remainder of the dataset. Streaming export operations must also account for failures that occur after the HTTP response has begun. Once response headers have been transmitted, conventional HTTP error responses cannot be returned. In practice, such failures are typically handled through detailed server-side logging and request correlation identifiers that allow export failures to be traced. Clients consuming export files may also detect incomplete downloads and treat truncated files as failed exports.

Another advantage of the streaming architecture is its predictable resource utilization. Because records are processed sequentially, the application server only maintains a small working set of data in memory. This allows the export mechanism to scale more effectively compared to buffered export implementations that accumulate large in-memory structures before returning a response.

Throughput control is another important operational consideration. In most deployments, streaming export pipelines rely on the natural flow control provided by TCP and the buffering behavior of the underlying servlet container. As the client reads data from the connection, the server continues writing serialized records. If the client is slower than the server, TCP backpressure naturally slows the transmission rate. Production systems may also impose limits on export duration or dataset size to prevent excessively long-running requests.

Finally, monitoring and logging should be incorporated into the export pipeline to track long-running export requests and detect failures during streaming. Observability mechanisms can help identify slow queries, abnormal export durations, or network interruptions that may impact the reliability of large downloads.

Overall, the streaming-based export design provides a practical and scalable solution for large transaction data exports in REST-based systems while maintaining stable resource usage across both application and database layers.

\section{Conclusion}

Exporting large transaction datasets is a common requirement in enterprise financial systems, yet traditional REST-based implementations often rely on buffered response models that construct entire export payloads in application memory before transmission. While straightforward to implement, this approach introduces significant limitations in terms of memory consumption, response latency, and system scalability when handling large datasets.

This paper presented a streaming-based export architecture that delivers transaction records incrementally as they are retrieved from the database. By integrating cursor-based database retrieval with HTTP response streaming, the proposed approach eliminates the need to buffer the complete dataset in memory. Records are processed sequentially and transmitted progressively to the client, allowing export operations to begin immediately and maintain stable resource utilization throughout the request lifecycle.

The architecture also supports multiple financial export formats through a modular serialization layer, enabling compatibility with widely used accounting and financial management tools. Combined with appropriate operational practices such as query optimization, connection management, and monitoring, the streaming design provides a practical and scalable solution for implementing large transaction export services in REST-based systems.

Future work may explore additional optimizations such as adaptive batching strategies, parallel export pipelines, or integration with distributed data processing platforms to further improve throughput for extremely large datasets.

Overall, the streaming export architecture demonstrates how incremental data processing and HTTP response streaming can significantly improve the efficiency, responsiveness, and scalability of large data export services in modern enterprise applications.

\section*{Disclaimer}

The views and opinions expressed in this paper are those of the author and do not necessarily reflect the official policy or position of Capital One. This work reflects the author's personal research and experience.

\bibliographystyle{IEEEtran}

\end{document}